\documentclass[sn-mathphys,Numbered]{sn-jnl}

\usepackage{graphicx}%
\usepackage{multirow}%
\usepackage{amsmath,amssymb,amsfonts}%
\usepackage{amsthm}%
\usepackage{mathrsfs}%
\usepackage[title]{appendix}%
\usepackage{xcolor}%
\usepackage{textcomp}%
\usepackage{manyfoot}%
\usepackage{booktabs}%
\usepackage{algorithm}%
\usepackage{algorithmicx}%
\usepackage{algpseudocode}%
\usepackage{listings}%
\usepackage{url}

\theoremstyle{thmstyleone}%
\theoremstyle{thmstyletwo}%

\theoremstyle{thmstylethree}%

\begin{document}

\title[Article Title]{Three-body periodic collisionless equal-mass free-fall orbits revisited}


\author*{Ivan Hristov$^{1}$,  Radoslava Hristova$^{1}$,  
Veljko Dmitra\v sinovi\' c$^{2}$,  Kiyotaka Tanikawa$^{3}$}

\affil{$^{1}$ Faculty of Mathematics and Informatics, Sofia University, 5 James Bourchier blvd.,
Sofia 1136, Bulgaria}
\affil{$^{2}$ Institute of Physics Belgrade, University of Belgrade, Pregrevica 118, Zemun, 
P.O.Box 57, 11080 Beograd, Serbia}

\affil{$^{3}$ National Astronomical Observatory of Japan,
Mitaka, Tokyo 181-8588, Japan}

\email{ivanh@fmi.uni-sofia.bg}  \email{dmitrasin@ipb.ac.rs}  \email{tanikawa.ky@nao.ac.jp}


\abstract{Li and Liao announced (2019) discovery of 313 periodic collisionless orbits' initial conditions (i.c.s), 30 of which have equal masses, and 18 of these 30 orbits have physical periods (scale-invariant periods)  $T^{*}=T|E|^{3/2}<80$. 
That work left a lot to be desired, however, both in terms of logical
consistency and of  numerical efficiency. We have conducted a new search for periodic free-fall orbits, limited to the equal-mass case. Our search produced 24,582  i.c.s of equal-mass periodic orbits with scale-invariant period $T^{*}<80$, corresponding to 12,409 distinct solutions, 236 of which are self-dual.}




\maketitle

\section{Introduction}\label{s:Introduction}

Free-fall orbits were perhaps the earliest periodic orbits to be studied in the general, i.e., equal- or almost-equal-mass 3-body problem: in the late 1960's and early 1970's, Szebehely \& Peters, Standish, and H\'enon \cite{Szebehely:1967,Standish:1970,Henon1974}. Their search was presumably motivated by Burrau's (1913) hypothesis of periodic Pythagorean 3-body motion. Also in the late 1960s Agekian \& Anosova \cite{Agekyan_Anosova:1967} started the studies of ``global'' properties, such as the ``lifetimes'' before ``ejection'' of free-fall orbits. 
Consequently, one might expect a large body of literature on the subject, which is not the case, rather some special cases of non-periodic colliding orbits have been studied \cite{Tanikawa:1995,Tanikawa:1999, Umehara:2000, Tanikawa:2008,Lehto:2008,Tanikawa:2015,Tanikawa:2019,Liao:2021}, presumably for their mathematical properties.
The free-fall orbits may yet prove to be of astronomical relevance, as the (observed) Saturn-Janus-Epimetheus and Sun-Earth-Cruithne systems belong to the same topological family.

After some preliminary work by Iasko and Orlov \cite{Iasko2014,Iasko2015},
it was only in 2019 that Li and Liao \cite{Liao:2019} searched Agekyan-Anosova's domain ${\cal D}$ for periodic orbits and found
313 initial conditions (i.c.s) for free-fall (or ``brake'') collisionless periodic orbits in the Newtonian three-body problem, 
30 of these 313 i.c.s
being for equal-mass case. This number of discovered and identified free-fall orbits does not compare well/favourably with thousands of other types of periodic 3-body orbits that have been found over the past decade \cite{Suvakov:2013, {Suvakov:2013b}, {gallery}, {Dmitrasinovic:2015}, Dmitrasinovic:2018,Liao:2017}. Why have so few periodic free-fall orbits been found thus far? Can one do better, and at what cost?

We emphasize here that these 313 were i.c.s, and not 313 distinct orbits, as every periodic free-fall orbit has two, usually distinct, sets of i.c.s. The 30 equal-mass i.c.s correspond to 28 distinct orbits, while 4 i.c.s contain 2 secondaries/``duals'', which are not (really) necessary to specify the orbit. In fact, the 26 secondary/dual of the rest i.c.s are easily obtained by following the time evolution of the primary i.c. through one half-period $T/2$ and finding the primary i.c. such that the triangle formed from the three bodies is similar to the triangle formed at $T/2$. This way we easily obtain additional 14 secondary i.c.s. The other 12 of 26 orbits are identified as self-dual(!), which means that the triangles at $t=0$ and $t=T/2$ are congruent. In this case the time evolution takes an i.c. into its  mirror image (plus possibly a permutation of mass labels), and then back again. These orbits have a beautiful  spatiotemporal symmetry  \cite{Montgomery:2023}.  They are not expected, as, thus far, the only examples of such self-dual free-fall orbits were certain isosceles orbits. The ratio $12/28$ correspond to the large number of $\sim 43 \%$ of the found self-dual orbits by Li and Liao.

\begin{figure}[t]
\centering
\includegraphics[scale=0.5]{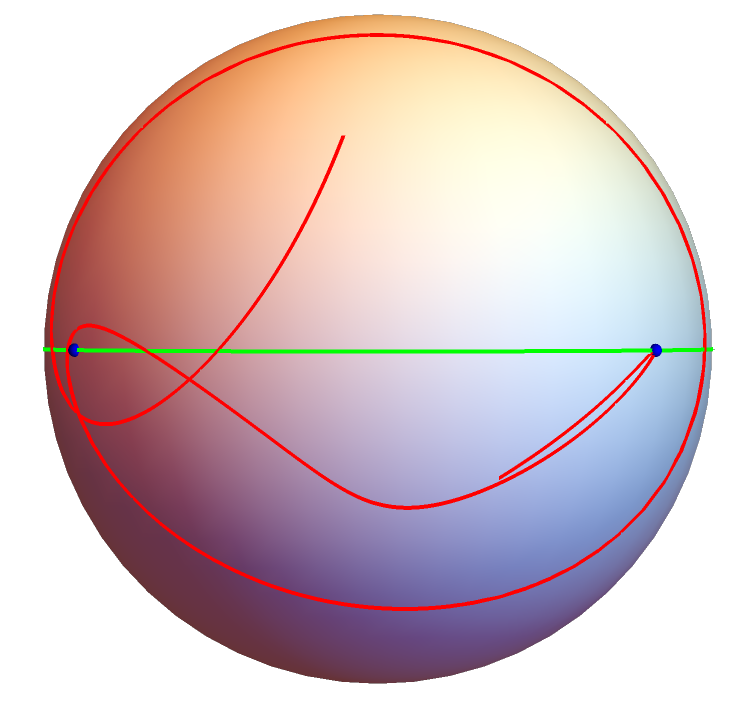}
\vspace{-0.2cm}
\caption{A free-fall  periodic orbit (solid red curve) on the shape-space sphere. Notice the two open ends, one above and another below the equator.}
\label{fig:brake_shape}
\end{figure}

There is another (mathematical) point of disagreement with Li and Liao \cite{Liao:2019}: they did not recognize the fact that free-fall periodic orbits form open segments on the shape sphere, (see Fig. \ref{fig:brake_shape}), rather than closed loops. Consequently, they used Montgomery's method \cite{Montgomery1998} of assigning two-symbol symbolic sequences to periodic orbits, which is not applicable here. Instead of Montgomery's method we use Tanikawa and Mikkola's syzygy counting method \cite{Tanikawa:2008,Tanikawa:2015} as a generator of symbolic sequences, and that match only for one half-period, as otherwise one would end up with trivial symbolic sequences. A special case of periodic free-fall orbits are the self-dual ones, which have particular structure of symbolic sequences, depending on the number of symbols in a half-period $T/2$.

Here we have re-visited this problem, using high-accuracy numerical methods and a high-performance computer (Nestum cluster, Sofia Tech Park, Bulgaria \cite{Nestum}). We have found:

1) 15,738 i.c.s corresponding to 12,409 distinct solutions with $T^*<80$. We use much  weaker restriction on the periods than Li and Liao: $T<\min(7,80/|E|^{3/2})$.  

2) When we apply time evolution of the (primary) i.c.s up to one half-period, we easily find 8,844 additional i.c.s.,
bringing the number of i.c.s up to 24,582. Some of the additional orbits are ones that are missed by the search, but others are with $T>7$ and $T^*<80$. 
Therefore by applying this procedure  we compensate to some extent that we use the less time consuming condition:  $T<\min(7,80/|E|^{3/2})$ instead of the ultimate one: $T<80/|E|^{3/2}$. Now only 236 solutions ($\sim2 \%$) are identified as self-dual, which contrast to the large number  $\sim 43 \%$ of Li and Liao. 

3) Certain partitioning of i.c.s within Agekyan-Anosova domain ${\cal D}$, which are in agreement with Tanikawa's division of ${\cal D}$ by $3$-cylinders and $4$-cylinders.

4) Distinct upper and lower bounds on the periods of orbits as (linear) functions of the length of symbolic sequence (word length).

As there is still an unknown number of
undetected periodic orbits, we are left with a significant uncertainty regarding the absolute number of periodic orbits with scale-invariant period smaller than some fixed value. The only thing that one can say with certainty is that the number of periodic free-fall orbits $N$ with a scale-invariant period $T^* = T|E|^{3/2}$ equal or shorter than any finite time $t$ is finite: $N(T^* < t) < \infty$. The scale-invariant period $T^* = T|E|^{3/2}$ of an orbit is bounded from above and below by bounds linearly proportional to the number of symbols $n$ in the symbolic sequence (``word'') describing the orbit. Thus, as the scale-invariant period $T^*$ grows, so does the number of possibly distinct periodic orbits, because the number $N_{n}$ of symbolic sequences (words) of length $n$ associated with them is bounded from above by $N_{n} \leq 2^{n/2}$. Thus, the only/best hope of a successful exhaustive search for periodic orbits is at small values of the scale-invariant period $T^*$.

The present paper falls into 5 sections.
After the present Introduction, in Sect. \ref{s:preliminaries} we present the preliminaries, such as the definition of Agekyan-Anosova's domain, of the shape sphere, and of the symbolic sequences used to describe three-body orbits. In Sect. \ref{s:Results} we present the results. In Sect. \ref{s:Conclusions} we summarize and draw conclusions.

\section{Preliminaries}
\label{s:preliminaries}

There are certain general features of three-body problem, as well as some features specific to the free-fall case, that need to be specified.

\subsection{Three-Body Jacobi Variables}
\label{ss:3bvbls}

Because the return proximity function is a function of twelve variables, it is difficult to systematically vary all twelve initial conditions to find a periodic solution. Therefore we eliminate all constants of the motion, and thus reduce the number of variables of the proximity function. 
One way to do this 
is by changing the three-body Cartesian variables to relative (Jacobi) ones 
and to the shape sphere. We shall use the latter for classifying periodic solutions. As most of these variables remain unchanged for arbitrary masses, we discuss the general case here.

The center-of-mass (CM) two-vector ${\bf R}_{\rm CM}$ is defined for arbitrary masses as
\begin{equation}
{\bf R}_{\rm CM} = \frac{m_1 {\bf r}_1 + m_2 {\bf r}_2 +
m_3 {\bf r}_3}{\sum_{i=1}^3 m_i}.
\end{equation}
${\bf R}_{\rm CM}$ is a constant of the motion if the total linear momentum
${\bf P} = {m_1 {\dot {\bf r}}_{1} + m_2 {\dot {\bf r}}_{2} +
m_3 {\dot {\bf r}}_{3}}$ equals zero. 
The 
three-body dynamics is
simplified by using the two relative coordinate vectors introduced by Carl Jacobi: 
\begin{equation}
{\bf \rho} = \frac{1}{\sqrt{2}}({\bf r}_1 - {\bf r}_2) \mbox{ and }
{\bf \lambda} = \frac{1}{\sqrt{6}}({\bf r}_1 + {\bf r}_2 -
2 {\bf r}_3),
\end{equation}
which are applicable even for unequal masses. We solve the equations of motion  
using Cartesian coordinates and then use ${\bf \rho}$ and ${\bf \lambda}$ to graphically represent the solutions.\footnote{The mass-weighted Jacobi vectors are not necessary for solving the equations of motion, nor do they represent the true geometry of the three-body trajectories. Consequently they can be avoided altogether.}
Thus we reduce the number of variables in the proximity function $d({\bf X}_0,T_0)$ from twelve to eight.

\subsection{Agekyan-Anosova's initial-condition space for the free-fall problem}
\label{ss:Definition}

As per definition of the free-fall orbit, there is at least one instance in time when all three bodies are at rest. We use that instance as our initial time $t_0 =0$, to define the initial conditions, which only concern the three bodies coordinates, i.e., the shape and the size of the triangle subtended by the three bodies. The historically accepted  convention, due to Agekyan and Anosova \cite{Agekyan_Anosova:1967}, see also Tanikawa et al. \cite{Tanikawa:1995,Tanikawa:2008,Tanikawa:2015}), for these variables is distinct from the shape sphere and the hyper-radius defined in \ref{ss:shape space}. That leads to certain complications, discussed below.

\begin{figure}
\centerline{\includegraphics[width=0.4\columnwidth,,keepaspectratio]{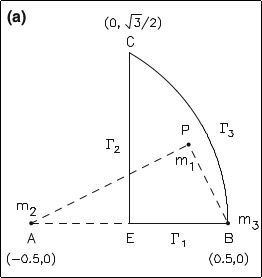}}
\caption{The initial position of three bodies in the configuration space. The initial positions of mass no. 2 and mass no. 3 are located at points $A(-0.5, 0)$ and $B(0.5, 0)$, respectively. The initial position of mass no. 1 is at the point $P(x, y)$ in the region ${\cal D}$.}
\label{fig:Agekyan_Anosova}
\end{figure}

The definition of the
initial-condition (i.c.) space for the free-fall 3-body problem was given by
Agekyan and Anosova. They put body number 2, with mass $m_2$ at $A(-0.5, 0)$ and body number 3 with mass $m_3$ at $B(0.5, 0)$ both on the $x$-axis of the $(x, y)$-plane. The body number 1 with mass $m_1$ is put at any place P within the circular segment
\begin{equation}
{\cal D} = \left\{ (x, y) : x \geq 0; y \geq 0; (x + 0.5)^2 + y^2 \leq 1 \right\} .
\label{e:Agekyan_Anosova}
\end{equation}
This is the Agekyan-Anosova domain ${\cal D}$, see Fig. \ref{fig:Agekyan_Anosova}.
As Tanikawa \cite{Tanikawa:1995} noted, this is not a unique choice, as can be seen by the fact that the vertical $x$ axis in Fig. \ref{fig:Agekyan_Anosova} describes isosceles triangles, but so does also the circular boundary of the Agekyan-Anosova domain ${\cal D}$.

\begin{figure}
\centerline{\includegraphics[width=0.4\columnwidth,,keepaspectratio]{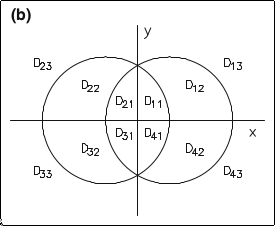}}
\caption{Tanikawa's 
triangle shape space. $D_{11}$ corresponds to the D-shaped domain in Fig. \ref{fig:Agekyan_Anosova}. The other $D_{ij}$ are obtained by reflections. For example, $D_{21}$ is the mirror image of $D_{11}$ with respect to the $y$-axis; $D_{12}$ is the mirror image of $D_{11}$ with respect to their boundary circle (cf. \cite{Tanikawa:2015}).}
\label{fig:Tanikawa_fig1}
\end{figure}

There are three other similar (in the sense that they can be obtained by reflections, about the $x$, and/or the $y$ axis, from this one) domains, another four compact and four infinite domains (12 {\it in toto}). They correspond to spatial reflections and different  permutations of mass labels (1,2,3) of the three bodies, see Fig. \ref{fig:Tanikawa_fig1}.

Permutation symmetry is perhaps most easily seen/illustrated on the shape sphere, see  Fig. \ref{fSFIG8} in Sect. \ref{ss:shape space}, where cyclic permutations correspond to rotations through $2\pi/3$ about the $z$-axis, whereas two-body permutations correspond to reflections about three Euler meridians. These meridians partition the shape sphere into six ``slices''/wedges, each one having a half ``north'' and ``south'' of the equator. These 12 sectors correspond to Tanikawa's 12 sectors in the $(x,y)$ plane, see Fig. \ref{fig:Tanikawa_fig1}.

An apparent advantage of the Agekyan-Anosova domain ${\cal D}$ over some of the other domains is that it is finite, but that may, in certain cases be an actual disadvantage, considering the fact that there are (probably) infinitely many periodic orbits located in a finite area. That means potentially high or even infinite density of periodic orbits, at least near the two-body collision points in the Agekyan-Anosova domain ${\cal D}$.

\subsection{The shape space of triangles}
\label{ss:shape space}

\begin{figure}[t]
\centering
\includegraphics[scale=0.2]{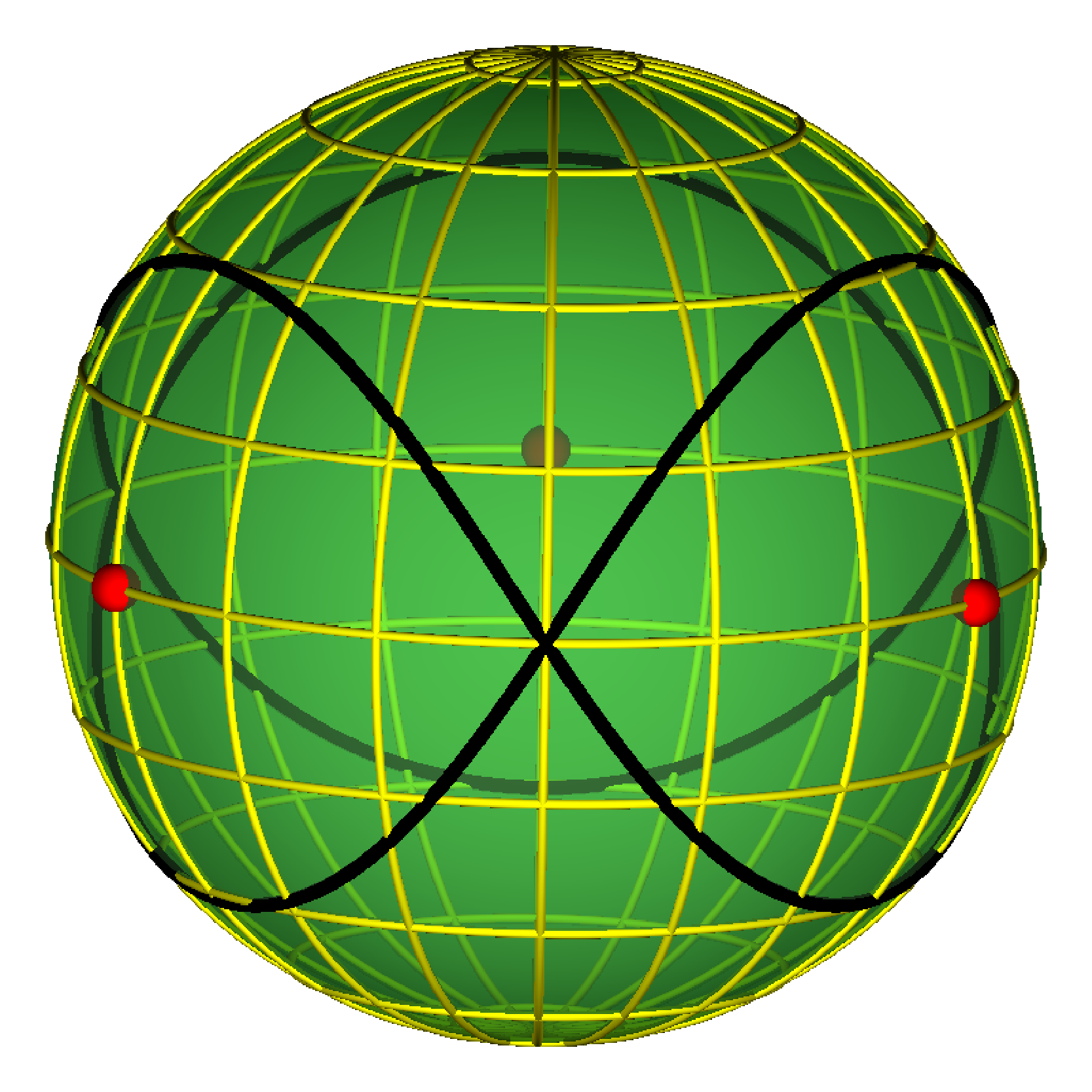}
\vspace{-0.2cm}
\caption{Figure-eight orbit (solid closed curve) on the shape-space sphere. Three two-body collision
points (red), singularities of the potential, lie on the equator.}
\label{fSFIG8}
\end{figure}

\begin{figure}[tbp]
\centerline{\includegraphics[width=2.5in,,keepaspectratio]{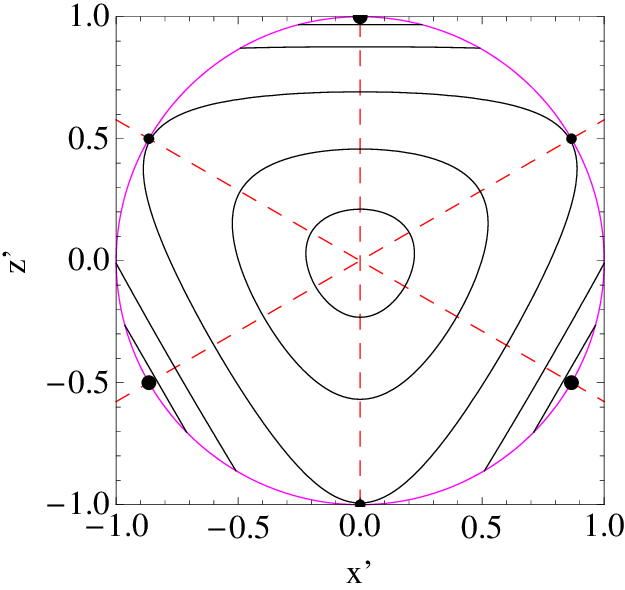}}
\caption{(Color online) Contour plot of the logarithm of the sum of Newton's two-body potentials for any fixed value of the hyper-radius $R$ in the equatorial plane of the shape sphere. The two straight dashed red  lines at angles of $\pm \frac{2 \pi}{3}$, and the vertical axis are the symmetry axes, i.e., the $s_2$ subgroups of the $s_3$ permutation group. The three collinear configurations in which one pair of particles has vanishing separation (``collision points'') are denoted by big solid circles, and the three collinear (``Euler points'') configurations in which one particle has equal separations from the other two are denoted by small solid circles. As one approaches the two-body collision points (three large solid points on the big circle), the equipotential
contour lines become increasingly dense,
finally reaching infinite density at these points, due to the singularities/poles present.}
\label{f:Newton contour}
\end{figure}

There are three independent scalar three-body variables:
$\lambda^2$, $\rho^2$, and ${ \rho} \cdot { \lambda}$.
The hyperradius
$R = \sqrt{\rho^{2} + \lambda^{2}}$ characterizes the overall size of the orbit and removes one of the three scalar variables. We may relate the three scalar variables to the unit three-vector $\hat {\bf n}$ defined by the
Cartesian components
\begin{equation}
\hat {\bf n} = \left (\frac{2 { \rho} \cdot { \lambda}}{R^2},
\frac{\lambda^2 - \rho^2}{R^2}, \frac{2 ({\rho} \times { \lambda})
\cdot { e}_z}{R^2} \right ).
\end{equation}
The domain of these three-body variables is a sphere with unit radius,
\cite{Montgomery1998}
as illustrated in Fig.~\ref{fSFIG8}.
The (shape) sphere coordinates depend only on the shape of the triangle formed by the three bodies, not on $R$ or on its orientation. The equatorial circle corresponds to collinear three-body configurations (degenerate triangles). The three points shown in Fig.~\ref{fSFIG8} correspond to two-body collisions, that is, singularities in the potential, see also Fig. \ref{f:Newton contour}.

Two angles parametrizing the shape sphere together with the hyperradius $R$ define the three-dimensional configuration space of the planar three-body problem.

The Smith-Iwai permutation-adapted 
(hyper)spherical angles ($\alpha$, $\phi$) are defined as follows
\begin{equation}
\hat {\bf n} = \left(\sin \alpha \cos \phi, \sin \alpha \sin \phi, \cos \alpha \right).
\label{e:n_Iwai2}
\end{equation}
which leads to
\begin{eqnarray}
\cos \alpha &=&
\left(\frac{2 { \rho} \times { \lambda}}{R^{2}}\right),
\label{e:alpha1} \\
\tan \phi &=& \left(\frac{n_x^{'}}{n_y^{'}}\right) =
\left(\frac{2 { \rho} \cdot { \lambda}}{{ \rho}^2 -
{ \lambda}^2}\right)
\label{e:phi1}.\
\end{eqnarray}
We see that the sign of $\cos \alpha$ changes every time the trajectory crosses the equator, or, equivalently, whenever the triangle passes through a syzygy (collinear  configuration).
Permutations are directly related to rotations about $z$-axis, i.e., to changes of the meridional angle $\phi$, see below.

Unlike many ``conventional'' periodic orbits, the free-fall orbits do not close a loop on the shape sphere, but rather form an open section of a curve, with two open ends, see Fig. \ref{fig:brake_shape}. Consequently, the topology of such a curve cannot be defined in the same way as for closed loops, see Sect. \ref{ss:symbolic_sequences}.

The natural domain of the permutation symmetric 3-body variables is a circle with unit radius, see Fig. \ref{f:Newton contour}. The points on the unit circle correspond to collinear configurations (``triangles" with zero area), or syzygies.

The permutation group on three objects $s_3$ consists of 6 elements, divided into 3 conjugacy classes, see textbooks, such as Stancu's \cite{Stancu:1996} \S 4 Permutation group $s_n$ and in Elliott and Dawber's \cite{Elliott_Dawber:1984} \S 17 Permutation group $s_n$.

The two straight lines at angles of $\pm \frac{2 \pi}{3}$, together with the vertical axis in Fig. \ref{f:Newton contour} are the three (reflection)
symmetry axes; these reflections correspond to the three ``two-body permutations"/transpositions in the $s_3$ permutation group. The two cyclic permutations of the $s_3$ permutation group correspond to the rotations through $\pm \frac{2 \pi}{3}$.

The six points where the symmetry axes cross the big circle in Fig. \ref{f:Newton contour} correspond to either a) three
collinear configurations (``shapes") in which one pair of particles has vanishing separation (big solid circles), i.e., ``sits on top of each other", or b) three collinear configurations (``shapes") in which one particle has equal separation from the other two, i.e. ``sits in the middle between the other two" (small solid circles). The center of the circle corresponds to the
equilateral triangle configuration (``shape''), which turns into a point when the hyper-radius $R \to 0$.

Note how the equatorial circle is divided into six identical ``pizza slices''/wedges, 
which number is doubled on the shape sphere, as, for each ``pizza slice'' in the equatorial plane there are two (equal-area) parts on the shape sphere: one ``north'' and one ``south'' of the equator, which are each other's mirror image.

\subsection{Syzygies as symbolic sequences for free-fall periodic orbits}
\label{ss:symbolic_sequences}

As stated in the Introduction, Sect. \ref{s:Introduction}, Montgomery's method \cite{Montgomery1998} of associating free-group elements does not work for periodic free-fall orbits, because they do not form a closed loop on the shape sphere. Rather, we use Tanikawa and Mikkola's method \cite{Tanikawa:2008,Tanikawa:2015} of associating a (finite) sequence  $w(0,T) = s_1 s_2 s_3 \ldots $ of three symbols $s_i$, say the numbers 1, 2 and 3 with each periodic orbit with period $T$. The three symbols/numbers (1,2,3)/ correspond to one of the three segments on the equator of the shape sphere wherein the orbit crosses the equator (which is guaranteed to happen by Montgomery's theorem 2007 \cite{Montgomery:2007}).

Of course, that still leaves us with an ambiguity regarding
permutations. Moreover, we shall only associate symbolic sequences with one half of the period $w(0,T/2)$, as the second half of a periodic orbit is represented by the (exact) inverse symbolic sequence $w(T/2,T) = w^{-1}(0,T/2)$, leaving a trivial (``unity'') sequence for (every) complete periodic orbit $w(0,T) = w(0,T/2)w(T/2,T) = 1$.

Each time a triple system becomes collinear, a symbol
is given according to the rule described at the top of this
subsection. Therefore, an orbit, except the one which terminates at triple collision, represented by an infinite continuous curve in the phase space is replaced by a bi-infinite (corresponding to the past and future) symbol sequence. We denote the symbols by
$s_i$ and the boundary between the past and future (zero time $t=0$ initial condition) by a period ($\bullet$), the present state being just after the period. Here we only consider the future symbol sequences. Then a (future) symbol sequence $s$ starting at the present can be written as
\[s = \bullet s_1 s_2 s_3 \ldots
~~~~~~~~~~~~(10)\]
Now, suppose that all the points in the initial condition
plane have their own symbol sequences (10), that is, the
orbits starting at points of the initial condition plane are
all integrated to the future. Of course, triple collision orbits have finite symbol sequences. If we truncate the symbol
sequences at the $n$-th digit, there are a finite number of possible combinations of symbols in these length-$n$ ``words''.
The point set of the initial condition plane whose symbol
sequences contain a particular initial word of length n
is called an ``$n$-cylinder''. For each $n$, a finite number of $n$-cylinders divide the initial condition plane into different sections.
Boundaries of the cylinders are formed with binary collision curves (BCCs) \cite{Tanikawa:2015, Tanikawa:2019}.

We use the shape sphere, as defined in \ref{ss:shape space} and in
Ref. \cite{Suvakov:2014}, to define the equator and syzygies.

\section{Results}
\label{s:Results}

\subsection{The return proximity function}
\label{ss:rpf}

The return proximity function $d({\bf X}_0,T_0)$ in phase space is defined as the absolute
minimum of the distance from the initial condition by
$d({\bf X}_0,T_0)=\min_{t \le T_0} \vert {\bf X}(t)-{\bf X}_0 \vert$,
where
\begin{equation}
\left \vert {\bf X}(t)-{\bf X}_0 \right \vert =
\sqrt{\sum_i^3 [{\bf r}_{i}(t) - {\bf r}_{i}(0)]^2 +
\sum_i^3 [{\bf p}_{i}(t)]^2 }
\end{equation}
is the distance (Euclidean norm) between two 12-vectors in phase space
(the Cartesian coordinates and velocities of all three bodies without
removing the center-of-mass motion). 
Searching for periodic solutions with a period $T$ smaller then a parameter $T_0$ is equivalent to finding zeros of the return proximity function. In this work, the  value of $T_0$ is taken to be $T_0(x,y)=\min(7,80/|E(x,y)|^{3/2})$. The numerical algorithm that we use for finding periodic orbits  is in principle those used in \cite{Liao:2017, Liao:2019}. This is the grid-search algorithm in combination with Newton's method \cite{Abad:2011}, where the computing of the coefficients of  the linear system at each step of Newton's method is done with the high order Taylor series method used with high precision floating point arithmetic \cite{Barrio:2011}. The obtained better efficiency from us can be explained with the use of a finer search-grid, the use of a modification of Newton's method with a larger domain of convergence \cite{Hristov:2021}, and some technical improvements. The numerical method will be given in much detail elsewhere.

\subsection{Distribution of i.c.s in Agekyan-Anosova's domain ${\cal D}$}
\label{ss:Disribution}

As a result of the search in Agekyan-Anosova's domain ${\cal D}$ with the restriction for the periods
$T(x,y)<\min(7,80/|E(x,y)|^{3/2})$,  we find 15,738 i.c.s corresponding to 12,409 distinct solutions.
9080 i.c.s turn out not to be in pairs (a distinct dual i.c. with the same $T^{*}$ does not exist). Simulating these 9,080 i.c.s up to $T/2$ and finding the i.c.s such that the triangle formed from the three bodies is similar to the triangle formed at $T/2$ (plus possibly a permutation of mass labels), we find  8,844 secondary i.c.s, bringing the number of i.c.s up to 24,582. This way we find together with some missed by the search i.c.s, also i.c.s for $T>7$ and $T^{*}<80$, i.e. this is a cheap way that easily compensate (to some extent) that 
we use the condition $T(x,y)<\min(7,80/|E(x,y)|^{3/2})$ instead of the ultimate but much more time consuming condition $T(x,y)<80/|E(x,y)|^{3/2}$. 
Additionally, this procedure is a way to identify the self-dual i.c.s, which means that the triangles at $t=0$ and $t=T/2$ are congruent and the time evolution takes an i.c. at $T/2$ into its  mirror image (plus possibly a permutation of mass labels), and then back again. We obtain $236=9,080-8,844$ such solutions. We will discuss them in \ref{ss:self_dual}.

\begin{figure}
\centerline{\includegraphics[width=0.7\columnwidth,,keepaspectratio]{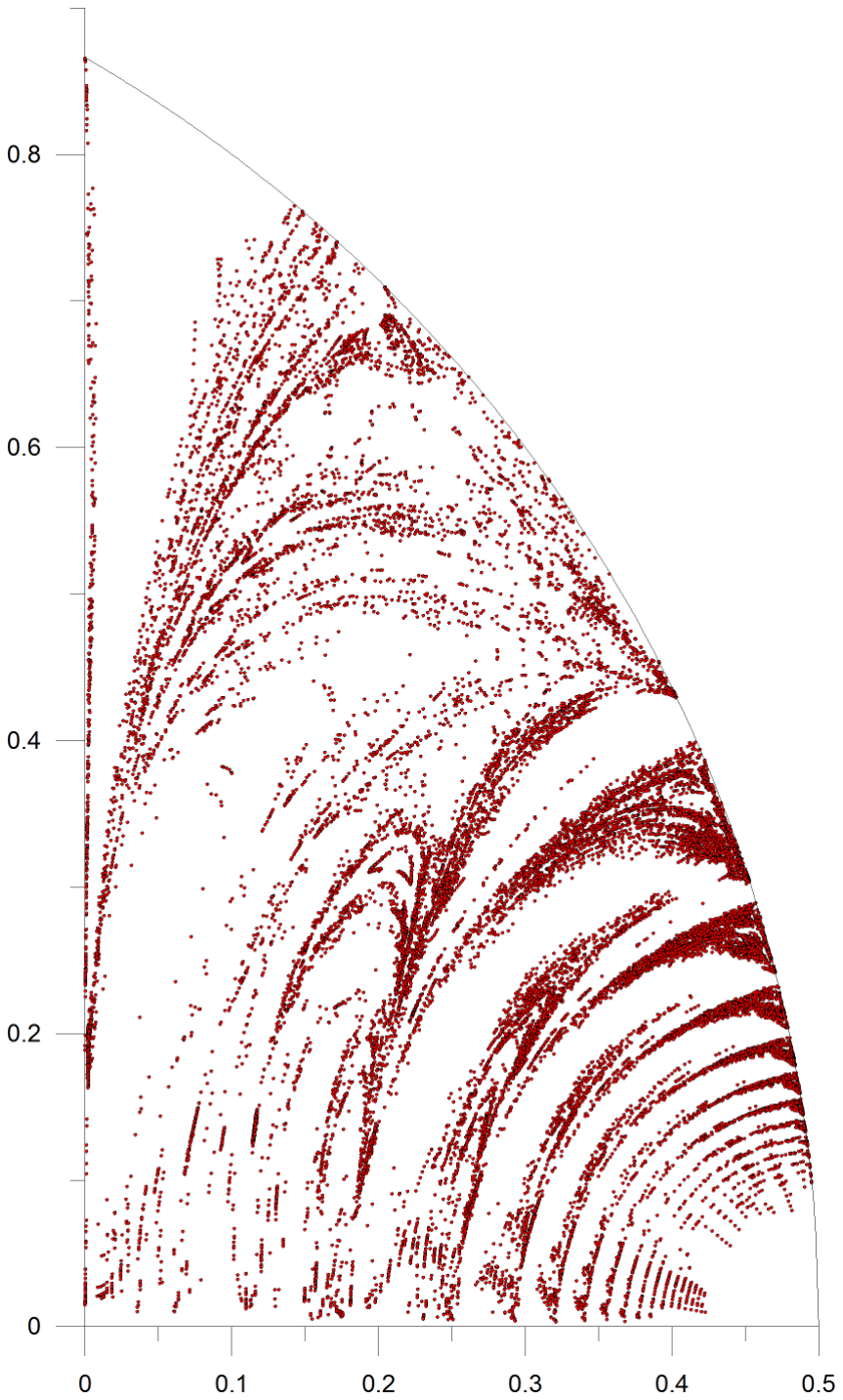}}
\caption{The 24,582 initial conditions of 12,409 free-fall orbits, including 236 self-dual ones.}
\label{fig:IC_last}
\end{figure}

All 24,582 i.c.s ordered by $T^{*}$ and given in the form of four numbers $(x,y,T, T^{*})$  with 80 correct digits can be downloaded from \cite{site}. All symbolic sequences (``words'') discussed in \ref{ss:symbolic_sequences} with their half-length $N/2$ and the plots in the real-space of 
some hundreds of the first orbits can be also found in \cite{site}. 

The distribution of i.c.s can be seen in Fig. \ref{fig:IC_last}. It is to be noted here that the structure of the picture is very similar to the pictures for ejection times and homology map in \cite{Lehto:2008} and also the picture for escape regions, binary collision curves, and triple collision points in \cite{Umehara:2000}. Particularly it is seen from \cite{Umehara:2000} that the distribution of i.c.s avoids the fast escape regions and is along the binary collision curves.

\begin{figure}
\centerline{\includegraphics[width=0.7\columnwidth,,keepaspectratio]{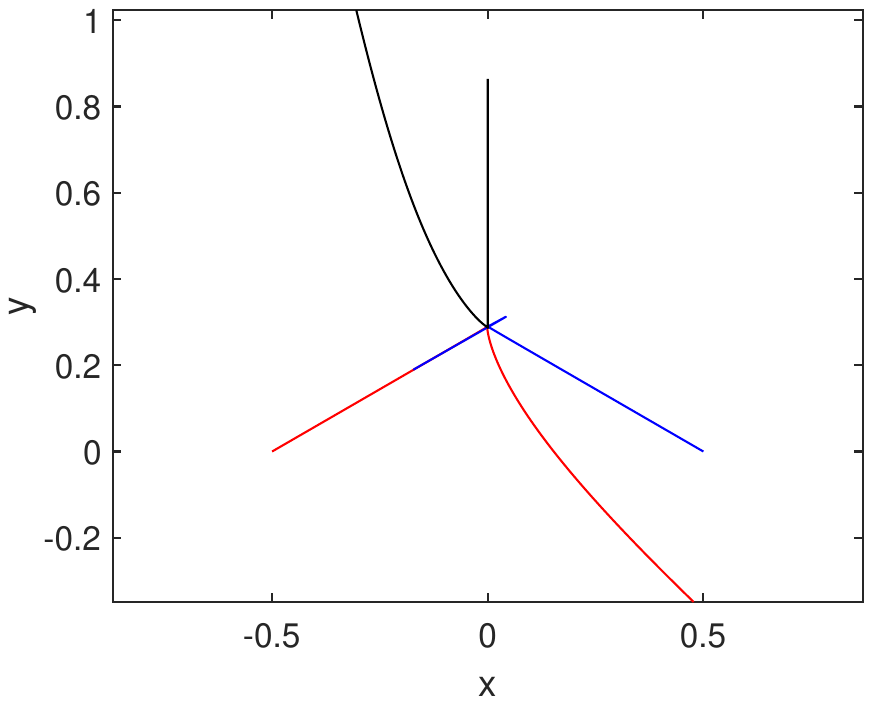}
\hskip-60pt\includegraphics[width=0.7\columnwidth,,keepaspectratio]{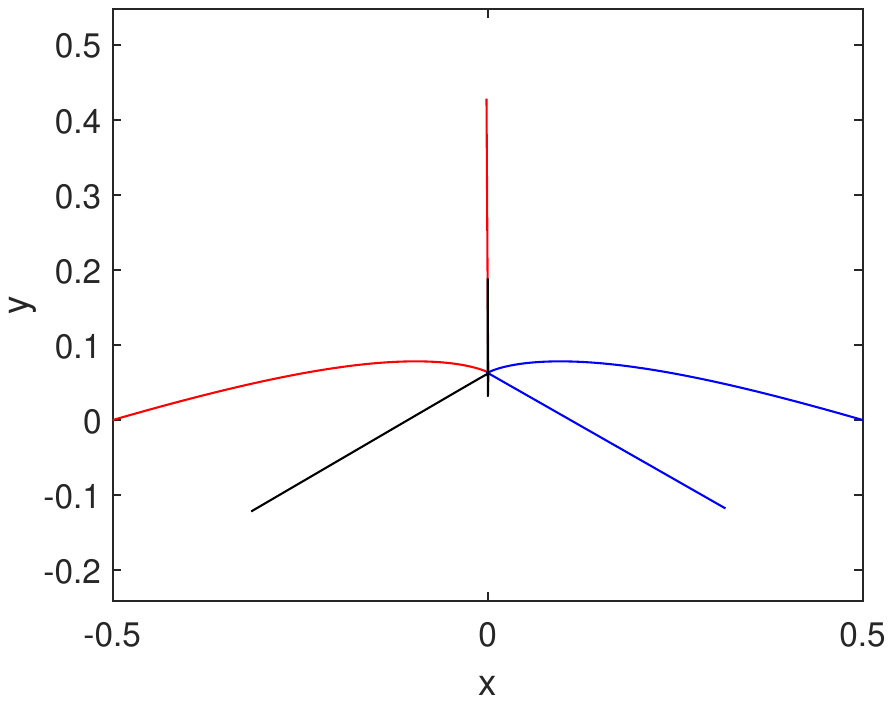}}
\vskip-80pt
\caption{The simplest ($N/2=5$) asymmetric orbit represented by two i.c.s, i.c. 1 (left), i.c.2 (right)}
\label{fig:sol1sol2}
\end{figure}

\begin{figure}
\centerline{\includegraphics[width=0.7\columnwidth,,keepaspectratio]{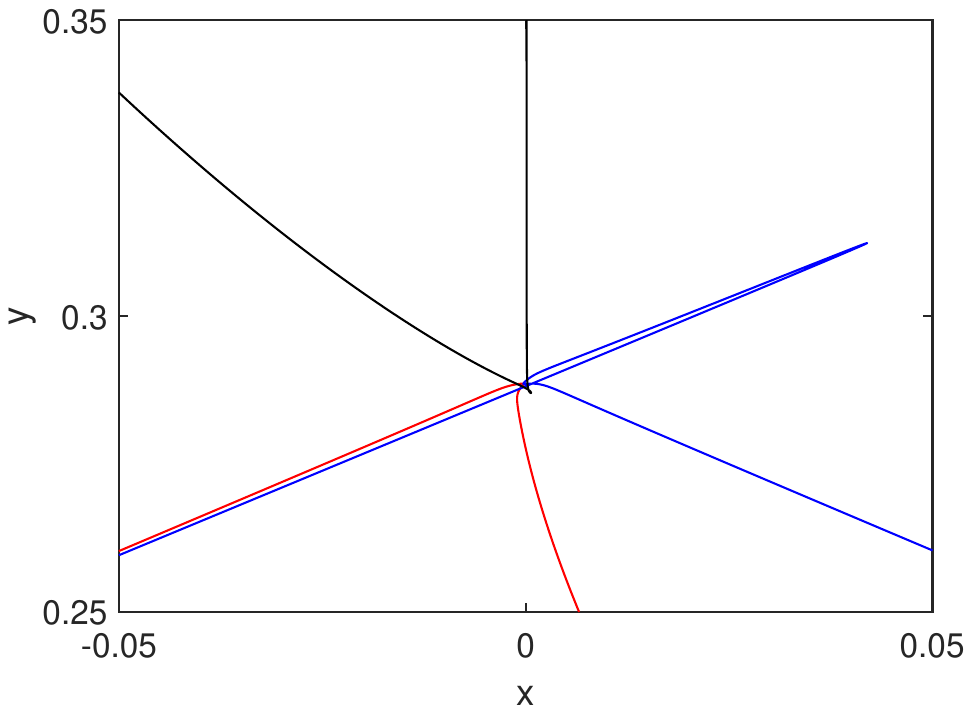}
\hskip-50pt\includegraphics[width=0.7\columnwidth,,keepaspectratio]{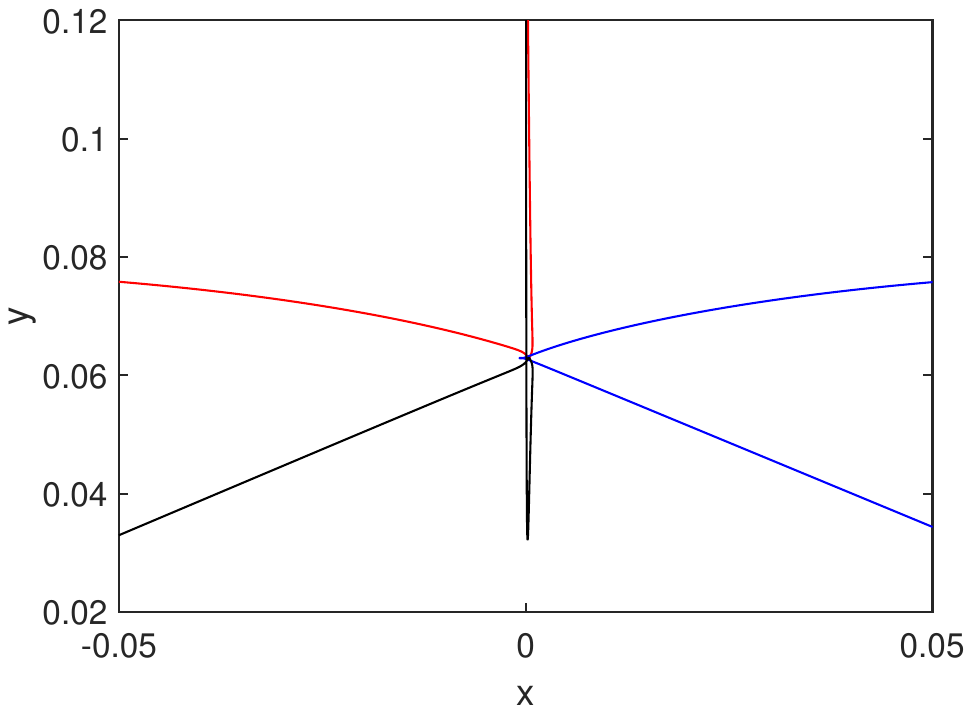}}
\vskip-80pt
\caption{The simplest ($N/2=5$) asymmetric orbit (zoomed), i.c. 1 (left), i.c.2 (right)}
\label{fig:solf1solf2}
\end{figure}

\begin{figure}
\centerline{\includegraphics[width=0.7\columnwidth,,keepaspectratio]{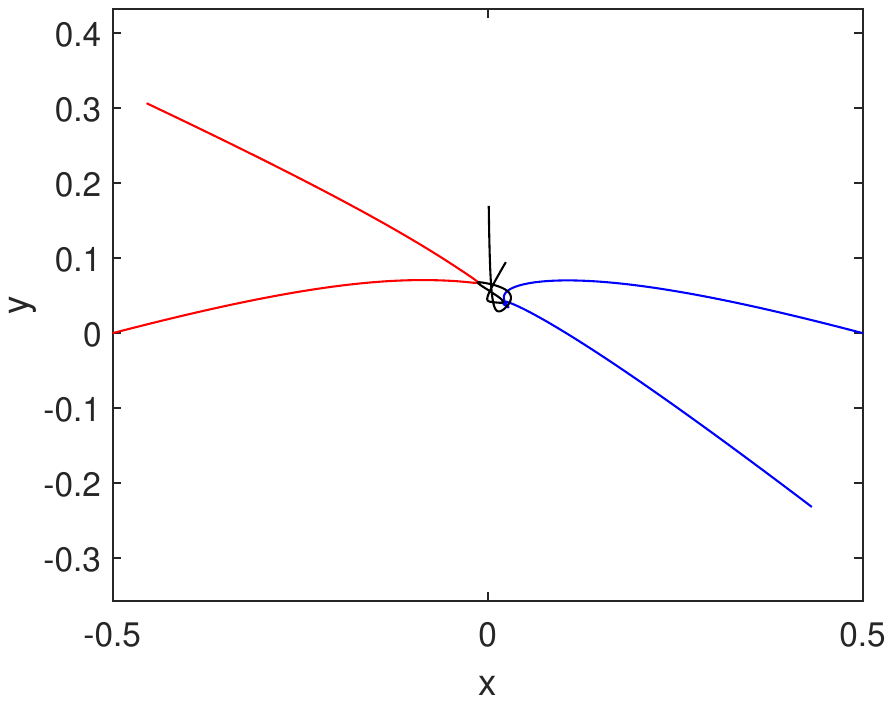}
\hskip-60pt\includegraphics[width=0.7\columnwidth,,keepaspectratio]{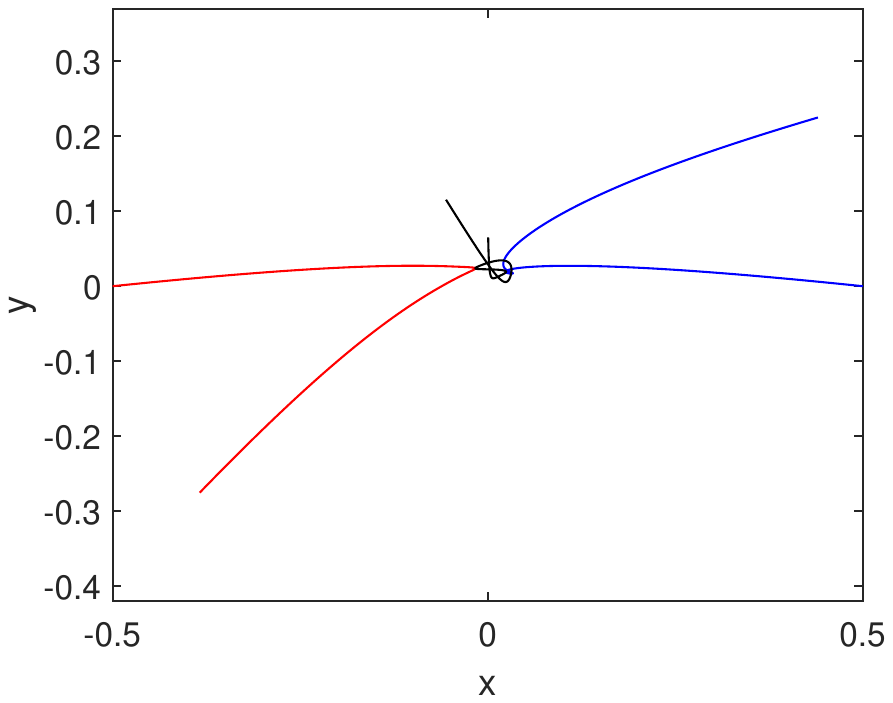}}
\vskip-80pt
\caption{The second simplest ($N/2=6$) asymmetric orbit represented by two i.c.s (i.c 3, i.c. 4)}
\label{fig:sol4sol5}
\end{figure}

The plots in the real-space for the simplest asymmetric solution (with $N/2=5$), represented by two distinct i.c.s, is given in Fig.\ref{fig:sol1sol2}.
The initial triangle for this solution is very closed to equilateral triangle, which leads to the Lagrange's triple collision orbit.
Although one can suppose a symmetry  in this solution by observing its plots, it is not the case, as seen from the zoomed versions of the plots in Fig. \ref{fig:solf1solf2}. The data $(x,y,T, T^{*})$ for this simplest solutions with 35 correct digits (approximately quadruple precision) can be seen in Table 1 (i.c.1 and i.c.2). Note that the scale-invariant periods $T^*$ for dual i.c.s are equal. The second simplest asymmetric solution is presented in Fig. \ref{fig:sol4sol5} 
(i.c.3 and i.c.4 in Table 1). 

 \begin{table}
 \begin{tabular}{ p{0.1cm} p{6cm} p{6cm} p{0.1cm}}
 \hline
 $i.c.$ & $\hspace{2cm}x$ & $\hspace{2cm}y$ & $\hspace{-0.2cm}N/2$\\
 {} & $\hspace{2cm}T$ & $\hspace{2cm}T^*$ & {}\\
  \hline
  1 &  0.10081002830904082427844218070523287e-3 & 0.86410262957299614417440390118167191e0  & 5\\
  {} &  0.27592315791656136782280882317399065e1 & 0.14361301848653203107026508191506541e2 & {}\\
  \hline
  2 &  0.20815094099721241496659892075009882e-4 & 0.18872009777437048186293267026882646e0  & 5\\
  {} &  0.13906231911063024951694149536454744e1 & 0.14361301848653203107026508191506541e2 & {}\\
  \hline
  3 &  0.13664936741948646596696276419442177e-2 & 0.16960184365485965711924464801930339e0  & 6\\
  {} &  0.14874034374340751927215339827790675e1 & 0.15583470416972024857327219803395329e2  & {}\\
  \hline
  4 &  0.61578935253763924971145321654327012e-3 & 0.64969497713782767052402395395661754e-1  & 6\\
  {} &  0.14078864983457596220282927174180473e1  & 0.15583470416972024857327219803395329e2  & {}\\
  \hline
\end{tabular}
{\caption {Data for the first 2 simplest asymmetric solutions (4 i.c.s) with 35 correct digits}}
\label{tab:table1}
\end{table}

\subsection{The role of binary collision curves in the Agekyan-Anosova's domain ${\cal D}$ }
\label{ss:bccs}
We insist on our orbits being collisionless, and this condition implies certain ``selection rules''. 
Tanikawa et al. have long studied binary collision curves (BCCs) for the Newtonian free-fall 3-body problem, mostly with regard to 3-body collisions and not with regard to periodic orbits. But, their results are equally valid and applicable to periodic orbits. The big picture is that the BCCs partition the Agekyan-Anosova domain into smaller compact subdomains, which form a pattern suspiciously similar to the pattern appearing in the pictures of ejection times and of the homology map in \cite{Lehto:2008}.

After computing the symbolic sequences of all found i.c.s, we partition (separate) them into different sets depending 
on the different starting 3 or 4 symbols. 
This way we want to match (to agree) the results for the initial symbols of found i.c.s with Tanikawa's division of ${\cal D}$ by $3$-cylinders and $4$-cylinders \cite{Tanikawa:2015, Tanikawa:2019}.

\begin{figure}
\centerline{\includegraphics[width=0.65\columnwidth,,keepaspectratio]{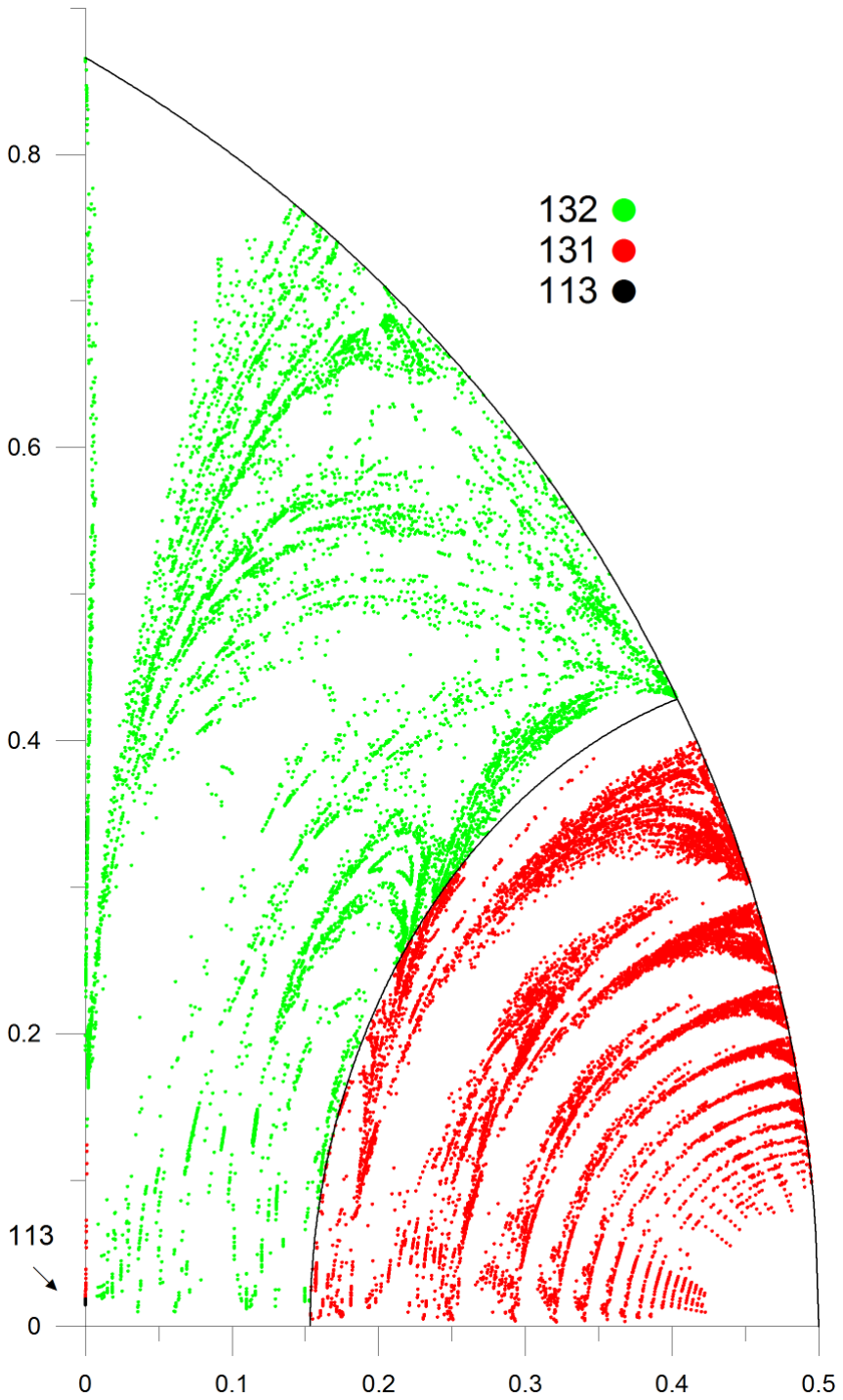}
\hskip-60pt\includegraphics[width=0.65\columnwidth,,keepaspectratio]{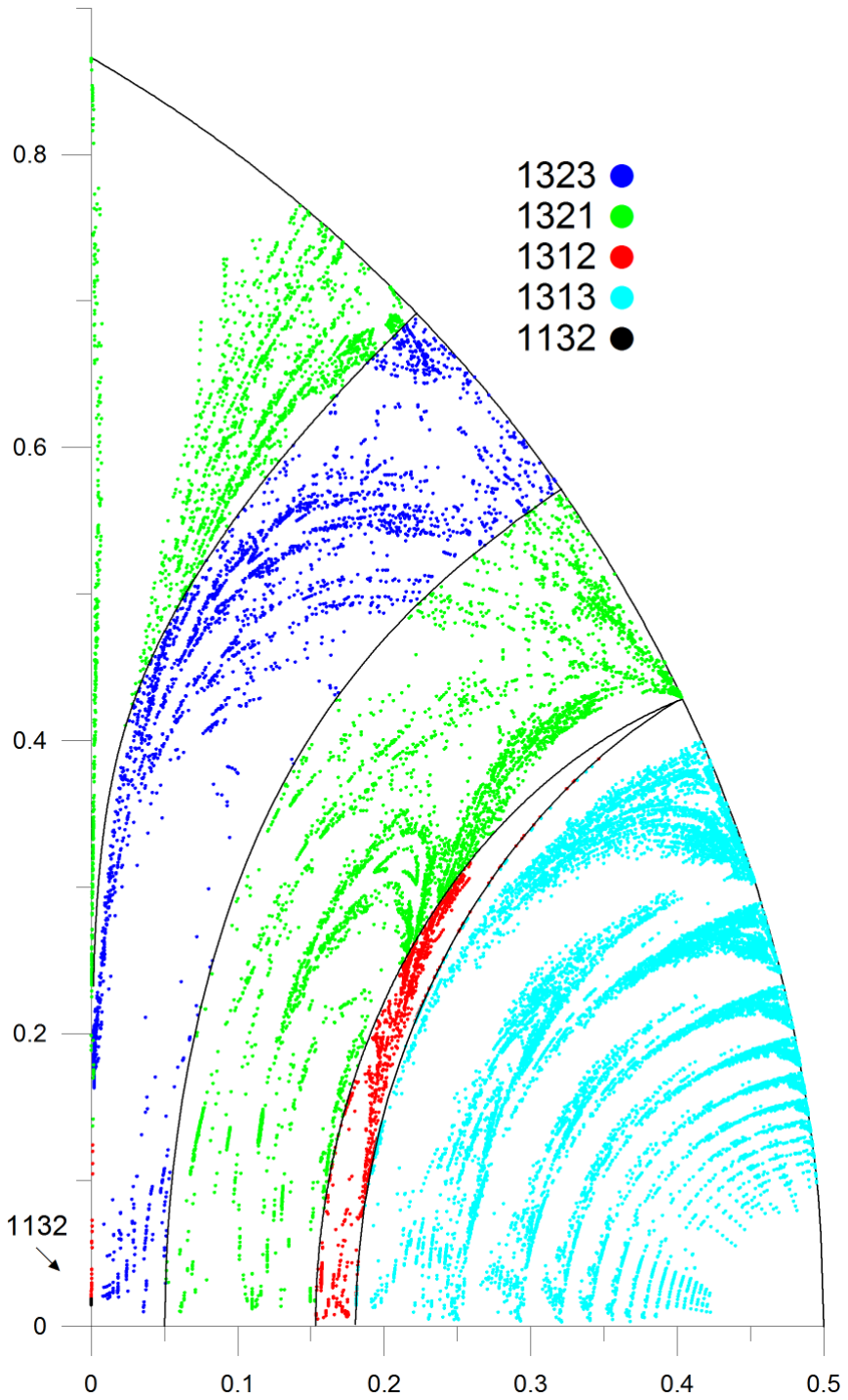}}
\caption{Partitioning of i.c.s in ${\cal D}$ by $3$-cylinders(left) and $4$-cylinders(right). Boundaries of the cylinders are formed with binary collision curves (BCCs) (the black curves inside ${\cal D}$)}
\label{fig:agree}
\end{figure}

In the case of considering the first three symbols we obtain three sets of i.c.s: starting with ``132'', ``131'', ``113''.  We color them in different colors - green, red, black, respectively. The binary collision curve (BCC) constituted the boundary of the 3-cylinders ``132'' and ``131'' from \cite{Tanikawa:2015, Tanikawa:2019} exactly separates the points with green and red color, meaning agreement (see Figure \ref{fig:agree} (left)). 
There is a small number of red and black points (``131'', ``113''-points) very close to the $y$-axis that make an exception
(they lie in the ``132'' - side of the BCC). This suggest that there is a structure of small scale close to the $y$-axis with
$|x|<0.01$. Computing  the BCCs in this region has not yet been done, as mentioned in \cite{Tanikawa:2019}, it will be considered elsewhere.

The case of starting 4 symbols is considered analogously. Now we have 5 sets of points: starting with ``1323'', ``1321'', ``1312'', ``1313'', ``1132''. We
color them in blue, green, red, cyan, black, respectively. The four binary collision curves (BCCs) constituted the boundary of the 4-cylinders from \cite{Tanikawa:2015, Tanikawa:2019} exactly separates the points with different colors (see Figure \ref{fig:agree} (right)), with the small exception mentioned above.

In the 4-cylinder 1132 close to the y-axis with $|x|<0.01$ (these are almost isosceles triangles close to the Euler point)
there are several examples of periodic orbits (13 i.c.s) that appear as the long awaited "stutter" orbits of \cite{Moeckel_Montgomery_Venturelli:2012}; we hope to return to this subject elsewhere.

\subsection{Self-dual (symmetric) orbits}
\label{ss:self_dual}
The self-dual orbits are those for which the triangles formed by the three bodies at $t=0$ and $t=T/2$ are congruent, i.e. dual i.c.s coincide.
We obtain 236 self-dual solution ($\sim 2\%$ from all 12,409 solutions), which contrast to the large number of $\sim 43\%$ (12 from 28) of Li an Liao.
This ambiguity can be explained by the  assumption that these solutions are generally easier to be caught numerically.
Also the percent of self-dual solutions among all solutions may depend on the considered value of $T^{*}$.

\begin{figure}
\centerline{\includegraphics[width=0.7\columnwidth,,keepaspectratio]{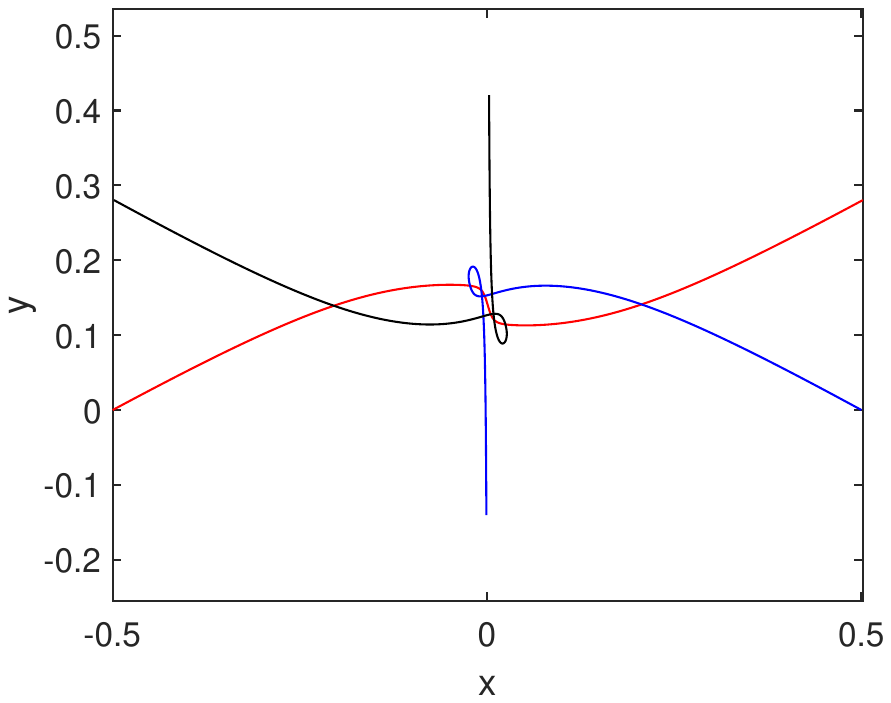}
\hskip-60pt\includegraphics[width=0.7\columnwidth,,keepaspectratio]{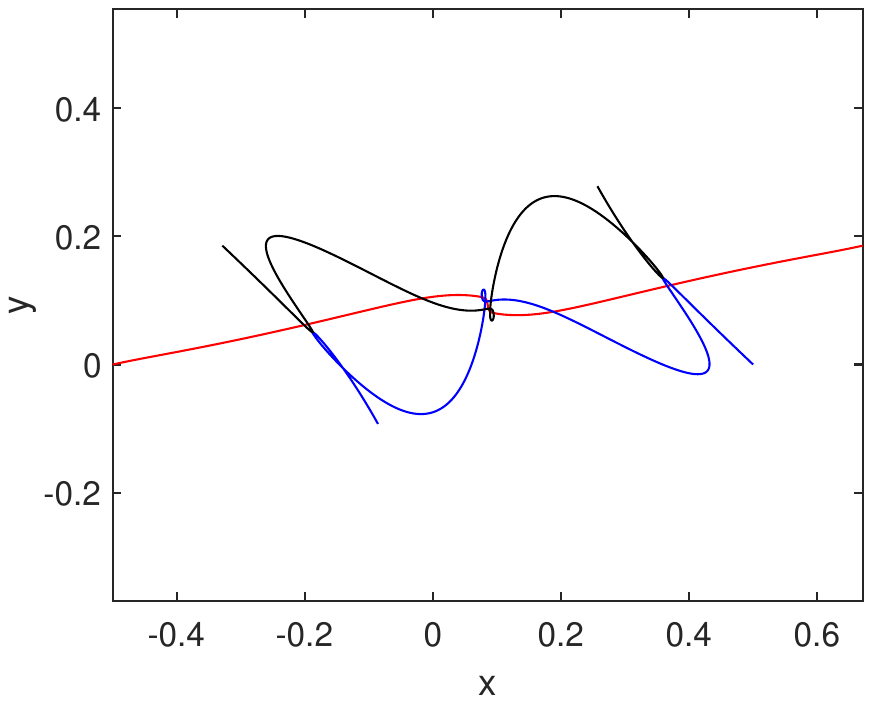}}
\vskip-80pt
\caption{The simplest free-fall symmetric orbits (i.c.s 1,2, $N/2=5,9$)}
\label{fig:sym1sym2}
\end{figure}

\begin{figure}
\centerline{\includegraphics[width=0.7\columnwidth,,keepaspectratio]{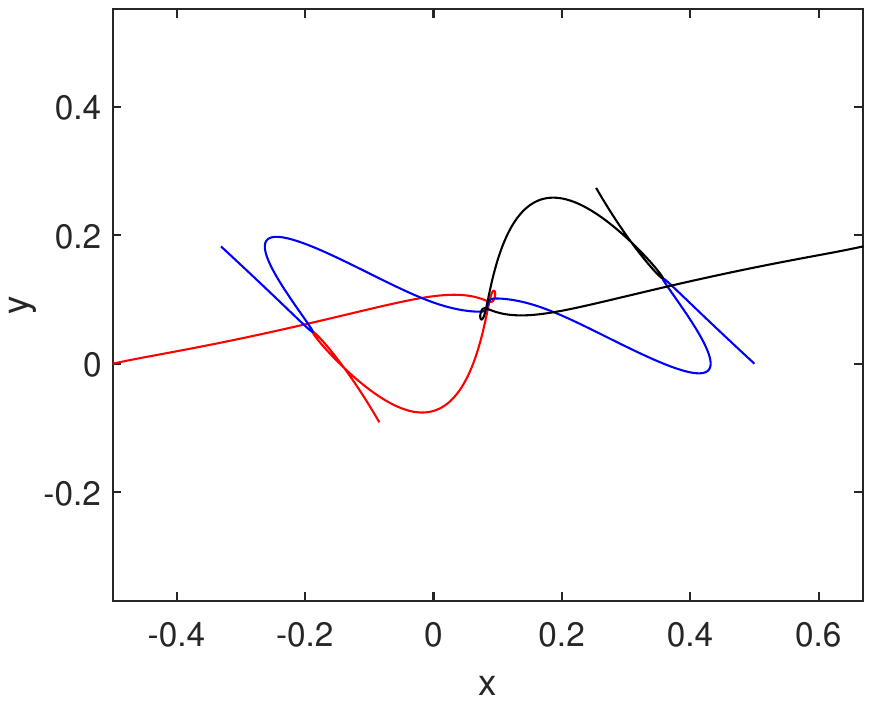}
\hskip-60pt\includegraphics[width=0.7\columnwidth,,keepaspectratio]{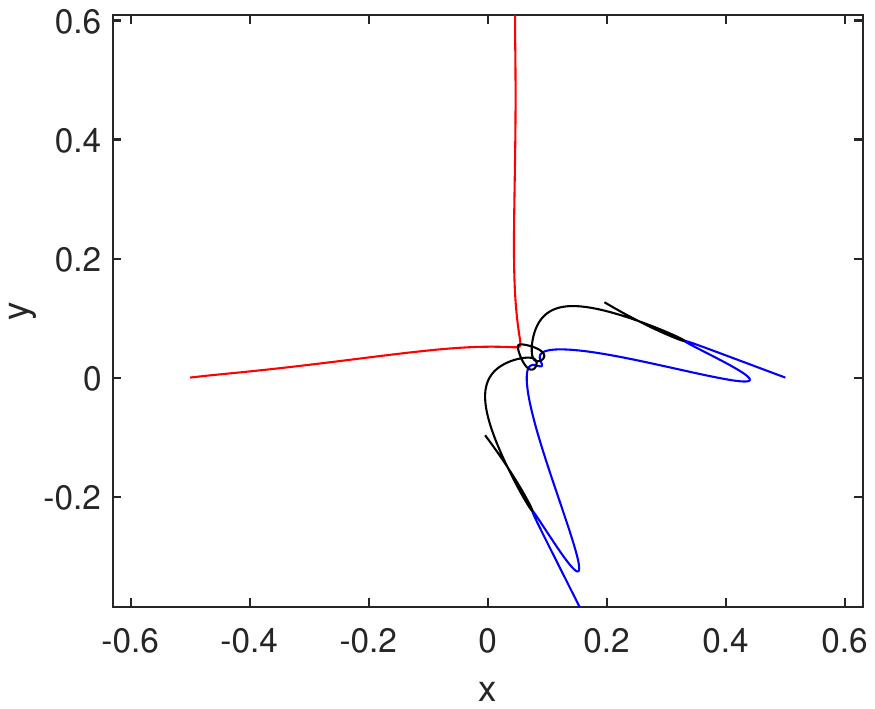}}
\vskip-80pt
\caption{The simplest free-fall symmetric orbits (i.c.s 3,4, $N/2=9,9$)}
\label{fig:sym3sym4}
\end{figure}

\begin{figure}
\centerline{\includegraphics[width=0.7\columnwidth,,keepaspectratio]{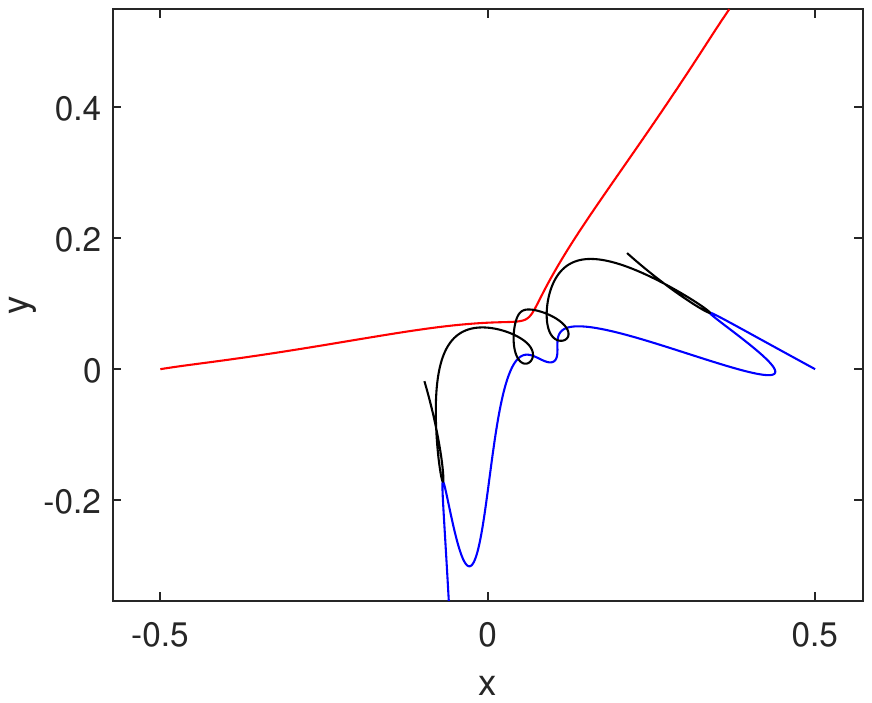}
\hskip-60pt\includegraphics[width=0.7\columnwidth,,keepaspectratio]{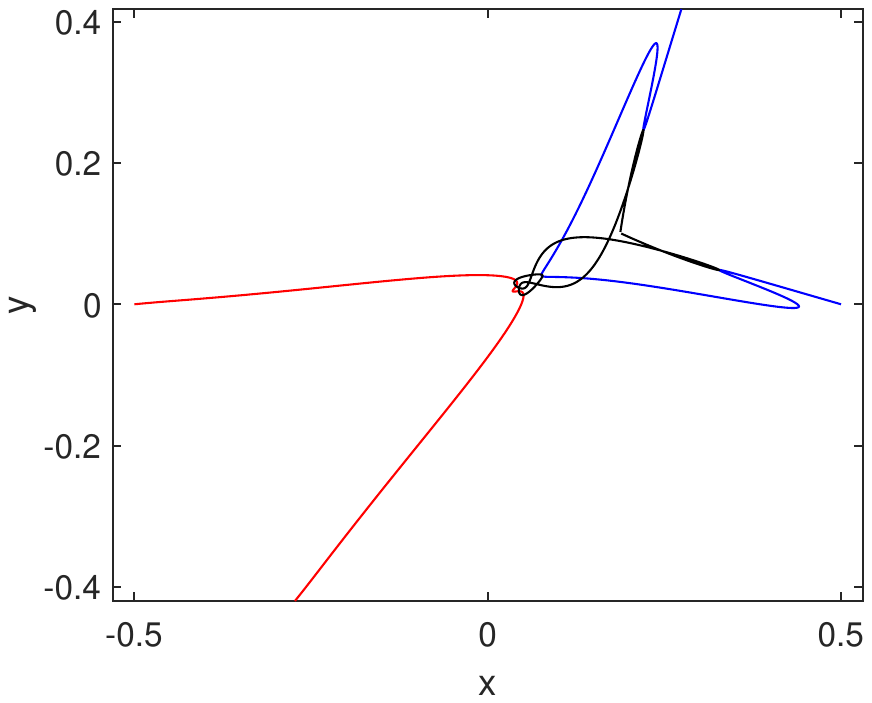}}
\vskip-80pt
\caption{The simplest free-fall symmetric orbits (i.c.s 5,6, $N/2=9,9$)}
\label{fig:sym5sym6}
\end{figure}

\begin{figure}
\centerline{\includegraphics[width=0.7\columnwidth,,keepaspectratio]{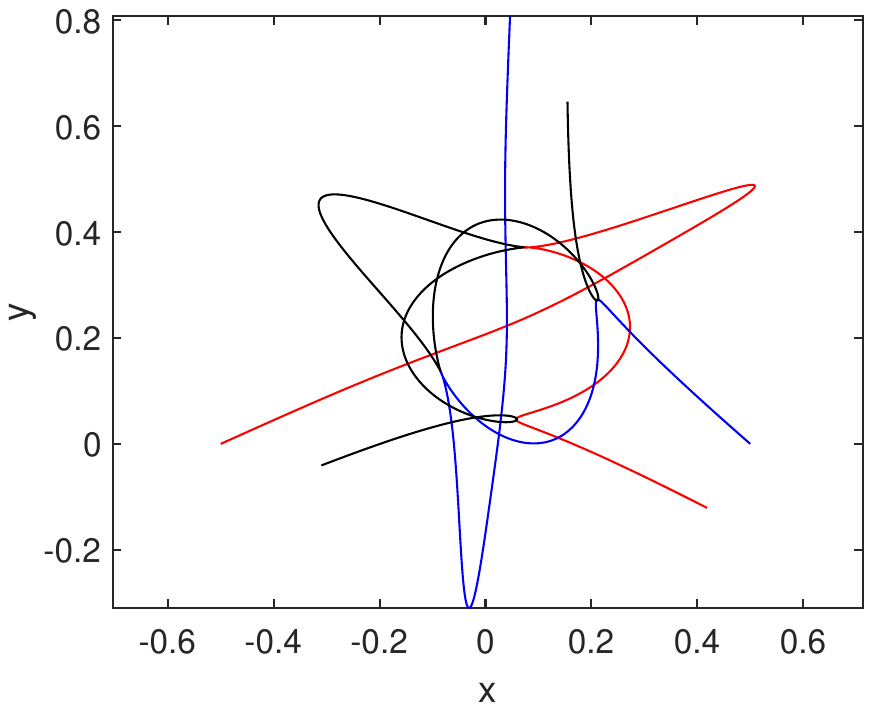}
\hskip-60pt\includegraphics[width=0.7\columnwidth,,keepaspectratio]{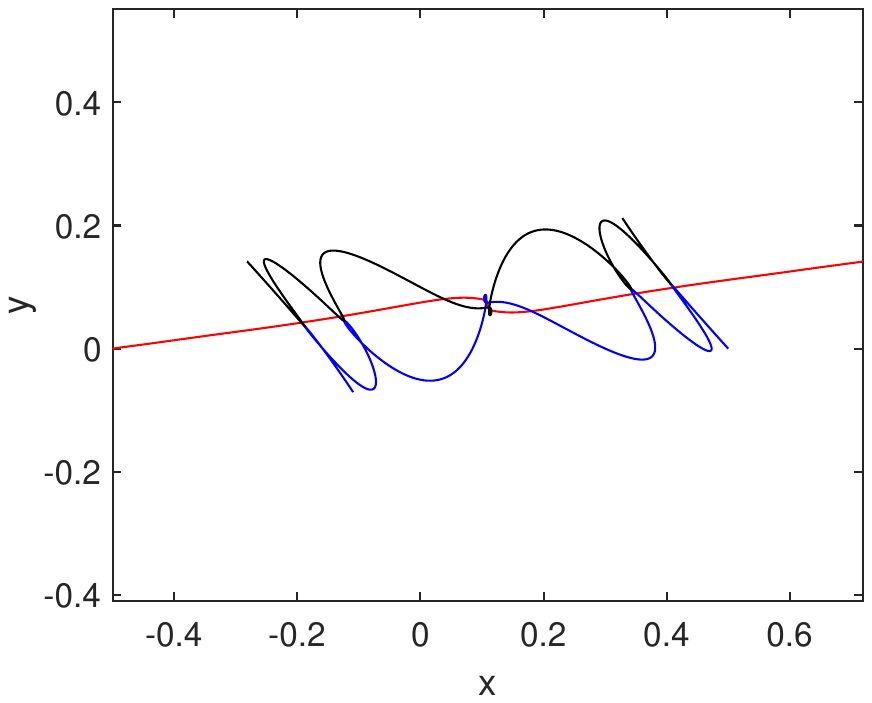}}
\vskip-80pt
\caption{The simplest free-fall symmetric orbits (i.c.s 7,8, $N/2=10,13$)}
\label{fig:sym7sym8}
\end{figure}

The beautiful  spatiotemporal symmetry of the self-dual solutions was for the first time discussed in \cite{Montgomery:2023}.
Here we present the 8 simplest ones, found by us. Their initial conditions with 35 correct digits is given in Table 2.
The corresponding plots in real-space are given in Figures \ref{fig:sym1sym2}, \ref{fig:sym3sym4}, \ref{fig:sym5sym6}, \ref{fig:sym7sym8}.
For all solutions, the map that gives on what vertex (mass label) each mass label in the initial triangle maps at $T/2$, is computed. For this map two possibilities are observed - or is an identity, or two indices are permuted. All self-dual solutions have the property that if we read the first half of the symbolic sequence from back to front and apply the map of vertices, the same sequence is obtained.
The symbolic sequences in their first half, their half-length $N/2$, the maps of the mass-labels and the permutations of mass-labels are given in Table 3.

According to the type of symmetry, the solutions are divided into two types.
The first type of symmetry is a reflection with respect to a line. 
In this case the initial triangle have to be flipped in order to match the second one in the plane.
These are solutions with i.c.s 4,5,6,7. In the case of this type of symmetry 
$N/2$ can be either even or odd. In the case of odd N/2 the map of the mass label is identity, and at T/4 the bodies lie on the axis of symmetry, in particular they are in syzygy (see i.c.s 4, 5, 6 for example). In the case of even N/2 two of the labels are always in permutation and in T/4 the body, which is not in permutation, lies on the axis of symmetry, and the other two do not lie, but are symmetrically located with respect to the axis (see i.c. 7 for example).

The second type of symmetry is a central symmetry with respect to a point. In this case the the triangles at t=0 and t=T/2 are obtained from each other with a rotation of 180 degrees. In this case $N/2$ is always odd and we always have a permutation of the labels of two of the bodies. At T/4, the bodies are in a syzygy, more precisely in ``Euler configuration''. The body in the middle is at the center of the symmetry and is the one that is not in permutation.

The observed properties of self-dual solutions are valid not only for here presented 8 of them, but for all.
The plots of all self-dual solutions and the maps of mass-labels can be downloaded from \cite{site}.

\begin{table}
 \begin{tabular}{ p{0.1cm} p{6cm} p{6cm} p{0.1cm}}
 \hline
 $i.c.$ & $\hspace{2cm}x$ & $\hspace{2cm}y$ & $\hspace{-0.2cm}N/2$\\
 {} & $\hspace{2cm}T$ & $\hspace{2cm}T^*$ & {}\\
    \hline
    1 &  0.26853656879186094505164950733431843e-2 & 0.42051392341346323788698432377477461e0 & 5\\  
    {} & 0.17803915330283836111234811629291376e1 & 0.14571737413553345363987962673353442e2 & {}\\
    \hline
    2 &  0.25694268581619406691269277542214527e0 & 0.27837466215631876401730693017725991e0 & 9\\
    {} & 0.21573615905021502937878732480336593e1 & 0.23729532401436922562439610275290658e2 & {}\\
    \hline
    3 &  0.25332955503585550078385245318739793e0 & 0.27395770000593360665609561689957488e0 & 9\\
    {} & 0.21486559741398094220093978785190743e1 & 0.23736057405684383328815469131959677e2 & {}\\
    \hline
    4 &  0.19608584373815407759186511872283212e0 & 0.12660087703751815111408497535170496e0 & 9\\
    {} & 0.19217736055334080320628254532331719e1 & 0.24456532037265669456189334514166059e2 & {}\\
    \hline
    5 &  0.21274288450316301296609635701735616e0 & 0.17736655380296837596730716436202892e0 & 9\\
    {} & 0.19915034464414970230624542166422446e1 & 0.24461630150799015077336406520614186e2 & {}\\
    \hline
    6 &  0.18896979078998075365597202763735798e0 & 0.10009595939299847925976966044966423e0 & 9\\
    {} & 0.18986434607202684769707963605898842e1 & 0.24469227124446530028878538869391828e2 & {}\\
    \hline
    7 &  0.15490107430442084270412195964824131e0 & 0.64608703728749114087928822042577474e0 & 10\\
    {} & 0.47578526770531448171326235767256789e1 & 0.30518409889739183020522928827494681e2 & {}\\
    \hline
    8 &  0.32792607131573432231865988797878982e0 & 0.21215354249540216269360745135067314e0 & 13\\
    {} & 0.22662686790481644958133858818197705e1 & 0.31908616428117479764156481776483866e2 & {}\\
    \hline
 \end{tabular}
 {\caption {Data for the first 8 simplest self-dual (symmetric) solutions with 35 correct digits}}
 \label{table2}
 \end{table}

  \vspace{30pt}

\begin{table}
 \begin{tabular}{ p{0.3cm} p{4cm} p{3cm} p{2cm} p{0.3cm}}
 \hline
  $i.c.$ & $symbolic \hspace{0.2cm} sequence \hspace{0.2cm} at \hspace{0.2cm} T/2$ & $maps \hspace{0.2cm} of \hspace{0.2cm} the \hspace{0.2cm}labels$ & $permutations$ & $N/2$\\

    \hline
    1 & 13213 & 1 $\rightarrow$ 3, 2  $\rightarrow$ 2, 3  $\rightarrow$ 1  &(13) & 5\\
    \hline 
    2 & 131321313 & 1  $\rightarrow$ 3, 2  $\rightarrow$ 2, 3  $\rightarrow$ 1 &(13) & 9\\
    \hline
    3 & 131231232 & 1  $\rightarrow$ 2, 2  $\rightarrow$ 1, 3  $\rightarrow$ 3 &(12) & 9\\
    \hline
    4 & 131323131 & 1 $\rightarrow$ 1, 2 $\rightarrow$ 2, 3 $\rightarrow$ 3  & e & 9\\
    \hline
    5 & 131323131 & 1 $\rightarrow$ 1, 2 $\rightarrow$ 2, 3 $\rightarrow$ 3  & e & 9\\
    \hline
    6 & 131232131 & 1 $\rightarrow$ 1, 2 $\rightarrow$ 2, 3 $\rightarrow$ 3  & e & 9\\
    \hline
    7 & 1321321321 & 1 $\rightarrow$ 1, 2 $\rightarrow$ 3, 3 $\rightarrow$ 2  & (23) & 10\\
    \hline
    8 & 1313132131313 & 1 $\rightarrow$ 3, 2 $\rightarrow$ 2, 3 $\rightarrow$ 1  & (13) & 13\\
    \hline
\end{tabular}

{\caption {The symbolic sequence, the maps of the mass-labels, the corresponding permutations and the half-length of the sequence for the first 8 simplest self-dual (symmetric) solutions}}
\end{table}

\subsection{Scale invariant period as function of symbolic sequence length}
\label{ss:T_N}

\begin{figure}
\centerline{\includegraphics[width=0.8\columnwidth,,keepaspectratio]{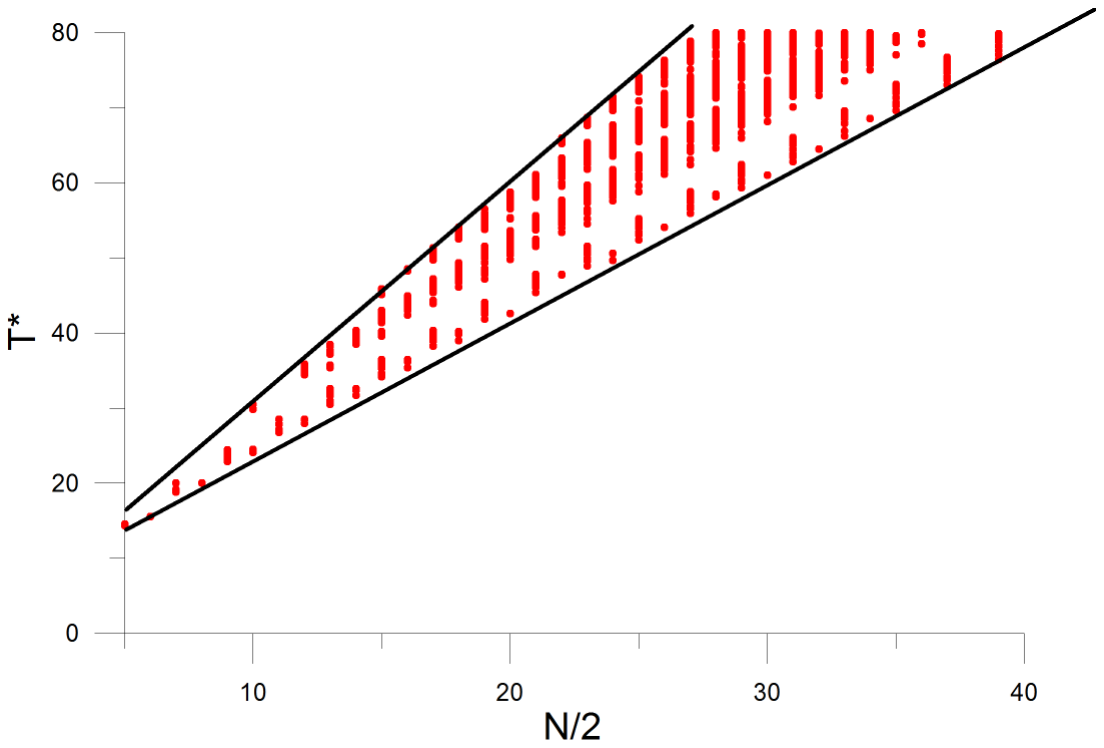}}
\vspace{-100pt}
\caption{$T^{*} = T|E|^{3/2}$
vs. $N/2$ for all
collisionless
free-fall orbits.}
\label{fig:T_N_all}
\end{figure}

The scale-invariant periods turn out to be bounded from above and from below by linear functions of symbolic sequence (``word'') length, see Fig. \ref{fig:T_N_all}. This linear dependence is in rough agreement with 
the linear dependence of closed-loop orbits first observed in \cite{Dmitrasinovic:2015} and later discussed in  \cite{Dmitrasinovic:2018},\cite{Dmitrasinovic:2017} with the distinction of a larger/wider spread of the data from a straight line. That spread can be understood in terms of the argument based on the analyticity/holomorphy of the action integral, as exposed in \cite{Dmitrasinovic:2018},\cite{Dmitrasinovic:2017}, however, with the following caveat: The curves on the shape sphere corresponding to free-fall orbits are not closed, but open, with two ends/terminals. A great circle can (always) be drawn through the two open ends, thus allowing two different closures of the curve, which, in turn, provide two bounds on the period.

\section{Concluding remarks and Outlook}
\label{s:Conclusions}
Now we summarize, conclude, and suggest future research.

1) The database of i.c.s for periodic collisionless equal-mass free-fall orbits is significantly expanded
from 30 to 25,582 and up to 80 significant digits.

2) The distribution of i.c.s in the Agekyan-Anosova's ${\cal D}$ domain shows a similar structure with many previously investigated 
properties of the problem in ${\cal D}$. Indeed there is a close correspondence with BCCs which correspondence ought to be explored further.

3) We found 236 self-dual (symmetric) solutions, which are only $\sim 2\%$ from all solutions.
   These 236 symmetric orbits ought to be of particular interest to mathematicians \cite{Montgomery:2023}.
   
4) We found several examples of periodic orbits in the 4-cylinder 1132 (almost isosceles triangles close to the Euler point)
that look like the long awaited "stutter" orbits of \cite{Moeckel_Montgomery_Venturelli:2012}. More about them elsewhere.

5) We obtain an agreement of computed symbolic sequences of the found i.c.s. with Tanikawa's division of Agekyan-Anosova's domain ${\cal D}$ by $3$-cylinders and $4$-cylinders. This correspondence ought to be extended to higher n-cylinders

6) The results suggest that the scale-invariant periods are bounded from above and from below by linear functions of symbolic sequence length,
   which is in agreement with previous results regarding closed-loop periodic orbits. This can/should be studied
   in greater detail.
   
7) Stability of orbits is of particular interest,  both because of the Birkhoff-Lewis theorem and of itself.


\bmhead{Acknowledgments}
The authors I.H. and R.H. acknowledged the access to the Nestum cluster @ HPC Laboratory, Research and Development and Innovation Consortium, Sofia Tech Park. V.D. thanks Mr. Bogdan Raoni\'c for his help in the early stages of this work. The work of V.D. was funded by the Serbian Ministry of Education Science and Technological Development, grant number 451-03-68/2020-14/200024.

\end{document}